\documentclass[aps,pre,twocolumn,superscriptaddress,floats,floatfix]{revtex4-2}
\usepackage{amssymb,amsmath,amsthm}
\usepackage{graphicx}
\usepackage{overpic}
\usepackage{epstopdf}
\usepackage{color}
\usepackage{subfigure}
\usepackage{epsfig}
\usepackage{rotating}
\usepackage{mwe}
\usepackage{float}
\usepackage{dcolumn,bm}
\usepackage{verbatim}
\usepackage{hyperref}
\usepackage{pgf,tikz}
\usepackage{cases}
\usepackage{afterpage}

\usetikzlibrary{calc}
\usetikzlibrary{arrows.meta, angles, quotes}
\usepackage{makecell}
\usetikzlibrary{3d}
\usepackage[normalem]{ulem}
\usepackage{xcolor}
\definecolor{Darkred}{RGB}{180, 0, 0}
\definecolor{Darkblue}{RGB}{26,68, 171}
\hypersetup{
  colorlinks = true,     
  urlcolor   = Darkblue, 
  linkcolor  = Darkblue, 
  citecolor  = Darkred   
}
\newcommand{\parheading}[1]{%
    \par\vspace{6pt}%
    \textit{#1}%
}

\begin{document}
\setlength{\parskip}{0pt}
\begin{abstract}
Axions and boson stars form as self-gravitating condensates of scalar-field dark matter. They are conventionally treated as separate cases based on the sign of their self-interaction, being attractive for axions and repulsive for bosons. Using simulations of binary mergers across the full self-interaction spectrum within the cubic-quintic Gross-Pitaevskii-Poisson framework, we show that the final mergers are governed by the single non-dimensional parameter $\Lambda$, which interpolates between the axion-like ($\Lambda<0$) and boson-like ($\Lambda>0$) limits. We show that the Gravitational-wave (GW) luminosity of the merger strongly depends on $\Lambda$, but the dominant chirp frequency $f_2$ follows a single scaling relation involving the compactness $\mathcal{C}$ of the mergers, independent of the sign of $\Lambda$. We further calculate the compactness analytically using the Gaussian ansatz for the mergers as a function of $\Lambda$, with good agreement with the compactness from simulations.
\end{abstract}

\title{Mergers of Self-Interacting compact objects: From Axions to Bosons}
\author{Diyanshu Wahane}
\email{diyanshus@iisc.ac.in}
\affiliation{Department of Physics, Indian Institute of Science, Bangalore 560012, India}
\author{Sanjay Shukla}
\email{s.shukla@tue.nl}
\affiliation{Department of Applied Physics and Science Education, Eindhoven University of Technology, 5600 MB Eindhoven, The Netherlands}
\author{Rahul Pandit$^1$}
\email{rahul@iisc.ac.in}

\maketitle


\parheading{Introduction:} Dark matter is an integral part of modern cosmology and astrophysics~\cite{bertone2018history}; yet, despite several decades of careful investigation, its nature remains elusive. The existence of dark matter has been inferred from various observations, including galactic rotation curves~\cite{rubin1978extended,bosma198121}, gravitational microlensing~\cite{aubourg1993evidence,vegetti2024strong}, large-scale galactic and super-galactic structures~\cite{frenk2012dark}, and anisotropies in the Cosmic Microwave Background (CMB)~\cite{aghanim2020planck}. Dark-matter searches~\cite{bozorgnia2024dark,boveia2018dark,roszkowski2018wimp} have targeted several candidates, including weakly interacting massive particles (WIMPs) and ultralight, weakly self-interacting bosons and axions~\cite{spergel2000evidence_CDM}, without conclusive evidence for any single candidate.

While such searches continue, it is important to study astrophysical objects that are candidates for dark-matter assemblies; these can be collapsed bosonic condensates~\cite{verma2021formation,verma2022rotating} or their axionic counterparts~\cite{braaten2019colloquium,Chavanis_att_PhysRevD.94.083007,shukla2024gravity,shukla2026self}. The axion, a pseudoscalar boson arising from the resolution of the strong-$\rm CP$ problem in QCD~\cite{peccei1977cp,weinberg1978the_axion,wilczek1978problem}, is a well-motivated dark-matter candidate~\cite{preskill1983cosmology}. QCD axions, even with their lowest theoretically predicted mass $\simeq 10^{-10} \mathrm{eV/c^2}$~\cite{di2021even}, cannot form galactic-scale ($\simeq$ kiloparsecs) dark-matter haloes; instead, they can form much smaller ($\simeq$ Km) compact objects, namely, \textit{axion stars}. A diffuse collection of such compact objects could conceivably constitute dark-matter haloes~\cite{barranco2013can}. The ultralight variant of the axion(mass $\lesssim 10^{-18}\  \mathrm{eV/c^2}$)~\cite{Hui2017ultralight}, on the other hand, forms a condensate at galactic scales. Such collapsed bosonic or axionic condensates are significant in dark-matter studies as they produce solitonic cores~\cite{schive2014cosmic,mocz2020galaxy} naturally; this is consistent with observed rotation curves~\cite{bosma198121}. Meanwhile, self-interactions between bosonic dark-matter particles help resolve~\cite{riotto2000if} the core-cusp problem~\cite{navarro1997universal,de2010core} in cold-dark-matter (CDM) haloes. This motivates treating self-interaction as a generic dark-matter property on top of their bosonic nature, while leaving open the question whether it is attractive, as for axions, or repulsive, as considered for bosonic dark matter~\cite{spergel2000evidence_CDM,braaten2016dense}.

Dark matter is observed primarily through its gravitational signatures~\cite{bertone2018history}. The GW190521 signal from the advanced LIGO-Virgo has even been interpreted as a possible vector-boson-star merger, as shown in the simulations of Ref.~\cite{bustillo2021gw190521}. Hence, if we wish to constrain its possible self-interactions, we should characterize gravitational waves (GW) that emerge from the mergers of compact dark-matter structures. Newtonian studies have investigated the merger dynamics of dilute axion stars~\cite{maseizik2024distributions, schwabe2016simulations}; and relativistic simulations have explored gravitational-wave emission from compact boson stars~\cite{chung2024multimessenger, siemonsen2023binary}. However, the GW phenomenology of axion-star mergers, within the Newtonian Gross–Pitaevskii–Poisson framework, remains uncharted. 
The dilute branch of axion stars~\cite{braaten2019colloquium, chavanis2018phase} are produced by the attractive quartic term in the instanton potential for axions~\cite{peccei1977CPconserv, marsh2016axion, chavanis2018phase} (for standard QCD axions, such stars have a maximum mass and must be far less massive than the solar mass~\cite{chavanis2023maximum}), whereas a stabilizing sextic repulsive term in the potential can prevent gravitational collapse. In contrast, boson stars are generally studied with a repulsive quartic interaction alone~\cite{palenzuela2007head, sivakumar2025revealing, shukla_PRF_2025,sivakumar2026vortex,shukla2026self}. 
Our study has been designed to carry out a systematic comparison of GW signatures of binary mergers by tuning a self-interaction parameter that allows us to go continuously from boson to axion stars.

We simulate binary mergers of compact objects, such as boson~\cite{verma2021formation,verma2022rotating,sivakumar2026vortex,shukla2026self} and axion~\cite{braaten2019colloquium,Chavanis_att_PhysRevD.94.083007,shukla2024gravity,shukla2026self} stars, using the quartic-sextic Gross-Pitaevskii-Poisson model; this allows us to examine the full range of quartic self-interactions, from repulsive to attractive. Thus, we can go from bosonic stars, with repulsive quartic intercations, to axionic stars, with attractive quartic interactions, by tuning the coefficient of the quartic term. The sextic term guarantees stability of the model for all signs of the quartic interaction term.     
Specifically, we use the interaction potential
\begin{equation}
V_{SI} = \frac{1}{2} g |\psi|^4+ \frac{1}{3} g_2|\psi|^6\,,
\label{eq:self-interaction}
\end{equation}
where the complex macroscopic wave function $\psi(\mathbf{r}, t)$ describes the bosonic or axionic condensates~\cite{sikivie2009bose, Hui2017ultralight,sivakumar2026vortex,shukla2026self}; $g$ and $g_2 > 0$ are the coefficients of the quartic and sextic terms; $g$ is positive for bosons and negative for axions~\cite{braaten2016dense,croon2019boson}.
We use the interaction~\eqref{eq:self-interaction} to look for relevant signatures in GW signals emitted by binary mergers of boson and axion stars [cf. Ref.~\cite{palenzuela2007head} for head-on collisions of boson stars]. In particular, we quantify the dependence of these signatures on the dimensionless self-interaction parameter
\begin{equation}
    \Lambda \equiv \frac{\tilde{g}}{\sqrt{\tilde{g}_2}}\,,
    \label{eq:Lambda}
\end{equation}
which measures the relative strengths of the quartic and sextic terms; clearly $\Lambda>0$ for boson stars $\Lambda<0$ for their axion counterparts; the tildes denote dimensionless interaction parameters [see Appendix~\ref{app:Dimensionless GPPE}].


Before we present the details of our study, we give a qualitative overview of our principal results. We focus on spiraling binary mergers as they lead to inspirals, mergers, and post-merger phases that are absent in head-on collisions. We use a continuous wavelet transform (CWT) to obtain the characteristic rising GW frequency signal, or \textit{chirp} signal, during an inspiral. We then demonstrate that the dominant merger frequency [see Ref.~\cite{bauswein2012measuring} and below] depends only on $\mathcal{C}$, the compactness of the final object, via a universal relation for compact objects~\cite{giudice2016hunting}. Using the minimization of the total energy of the system, we obtain analytically the relation $\mathcal{C}(\Lambda)$
between the  compactness and $\Lambda$; we then show that this relation is in consonance with our simulation results. Furthermore, we find that the GW luminosity $\mathcal{L}$ depends significantly on $\Lambda$: For axion-star mergers ($\Lambda < 0$) we demonstrate that $\mathcal{L}$ first increases sharply with time
and then decays after a prominent peak; in contrast, $\mathcal{L}$ increases and then approaches a large-$t$ asymptote for boson stars, especially if $\Lambda \gg 0$; i.e., at saturation we have a GW-power-emitting remanent. We finally comment on the implications of our results for current detectors.




\parheading{Gross-Piaevskii-Poisson framework:} At low temperatures, where axions are non-relativistic~\cite{sikivie2009bose}, axionic dark matter can be described by a complex macroscopic wave function $\psi(\mathbf{r}, t)$~\cite{sikivie2009bose, Hui2017ultralight}. In the low-energy regime, axions are described by the non-relativistic effective field theory~\cite{braaten2016nonrelativistic, Salehian2021}, where the full Einstein–Klein–Gordon system reduces to the Gross-Pitaevskii–Poisson equations at leading order. The Lagrangian of such an axionic system in the presence of a self-gravitational potential $\Phi$ is given by
\begin{eqnarray}
    \mathcal{L} &=& \frac{i\hbar}{2} \left( \psi^* \frac{\partial \psi}{\partial t}-  \frac{\partial \psi^*}{\partial t}\psi\right) -\frac{\hbar^2}{2m}|\nabla \psi|^2 - V_{SI} -\mu |\psi|^2\nonumber\\
    &&-m\Phi|\psi|^2- \frac{1}{8\pi G} (\nabla \Phi)^2\,,
    \label{eq:lagrangian}
\end{eqnarray}
where $m$ is the axion mass, $\mu$ the chemical potential, and $V_{SI}$ the self-interaction potential. The widely used phenomenological potential for the axion, motivated by non-perturbative instanton (superscript $i$) effects in QCD, is~\cite{peccei1977CPconserv, marsh2016axion, chavanis2018phase}
\begin{eqnarray}
    V_{SI}^{i} = \frac{m^2cf^2}{\hbar^3}\bigg[1-\cos\bigg(\frac{\sqrt{\hbar c}\phi}{f}\bigg) \bigg]\,,
    \label{eq:instanton}
\end{eqnarray}
where $c$ is the speed of light, $f$ is the axion-decay constant, and $\phi$ is the real scalar field that is related to the complex field $\psi$ via $\phi = \hbar/(\sqrt{2m})[\psi e^{-imc^2t/\hbar}+\psi^* e^{imc^2t/\hbar}]$. If we expand $ V_{SI}^{i}$ in powers of $|\psi|^2$ and truncate the expansion at order $|\psi|^6$ (required for stability if $g < 0$) we obtain the quartic-sextic interaction $V_{SI}$ [Eq.~\eqref{eq:self-interaction}]
with $g=\tfrac{4\pi a\hbar^2}{m}$ and $g_2=\tfrac{32 \pi^2\hbar^4a^2}{3m^3c^2}$, where the scattering length $a \equiv -(\hbar c^3m)/(32\pi f^2)$ is negative, because of the attractive quartic self-interactions between axions. For $g_2=0$, the system is extremely unstable, above a very low critical mass~\cite{chavanis2023maximum}, beyond which it collapses. For bosonic particles the scattering length $a>0$, so the sextic term is not required for stability; we retain it to describe attractive, intermediate, and repulsive cases using the single dimensionless parameter $\Lambda$. For the pure bosonic case, we compare the results with $g_2=0$ and $g>0$; this is the limit $\Lambda\rightarrow \infty$.
[For axion stars, which use a phenomenological instanton potential, the coefficients $\tilde{g}$ and $\tilde{g}_2$ are not independent but related; so our study explores $\Lambda$ as an independent phenomenological control parameter, moving away from axion-like towards purely repulsive boson-like values.]

The functional derivative of Eq.~\eqref{eq:lagrangian} gives the following cubic-quintic Gross-Pitaevskii-Poisson equation (cq-GPPE):
\begin{eqnarray}
i\hbar \frac{\partial }{\partial t}\psi(\mathbf{r}, t) &=& \bigg( -\frac{\hbar^2}{2m} \nabla^2 + g |\psi(\mathbf{r}, t)|^2 + g_2 |\psi(\mathbf{r}, t)|^4 \nonumber\\
&&+ m\Phi \bigg) \psi(\mathbf{r}, t)\,,
\label{eq:GPP_equation}
\\
\nabla^2\Phi &=& 4\pi Gm(|\psi|^2-\langle |\psi|^2\rangle)\,.\nonumber
\end{eqnarray}
We note that setting $g_2 = 0$ ($g>0$) recovers a conventional repulsive boson star~\cite{shukla2026self}; setting $g<0$ (with fixed $g_2>0$) gives the axion limit; retaining both $g>0 , g_2>0$ gives us what we call a \textit{q-boson} (where \textit{q}
denotes quintic). If we linearize Eqs.~\eqref{eq:GPP_equation} around a uniform background, we obtain the Jeans instability [see Appendix~\ref{app:GPPE}]; note, moreover, that the Eqs.~\eqref{eq:GPP_equation} conserve the number of particles $N$ and the total energy $E$.

We solve the Eqs.~\eqref{eq:GPP_equation} using a pseudospectral DNS~\cite{canuto2012spectral,verma2021formation,verma2022rotating,shukla2024gravity,shukla2026self} that we describe in the Appendix~\ref{app:model}, where we also give the initial conditions, and the units and dimensions.

\begin{figure*}[htb!]
   \centering
   \includegraphics[scale=0.17]{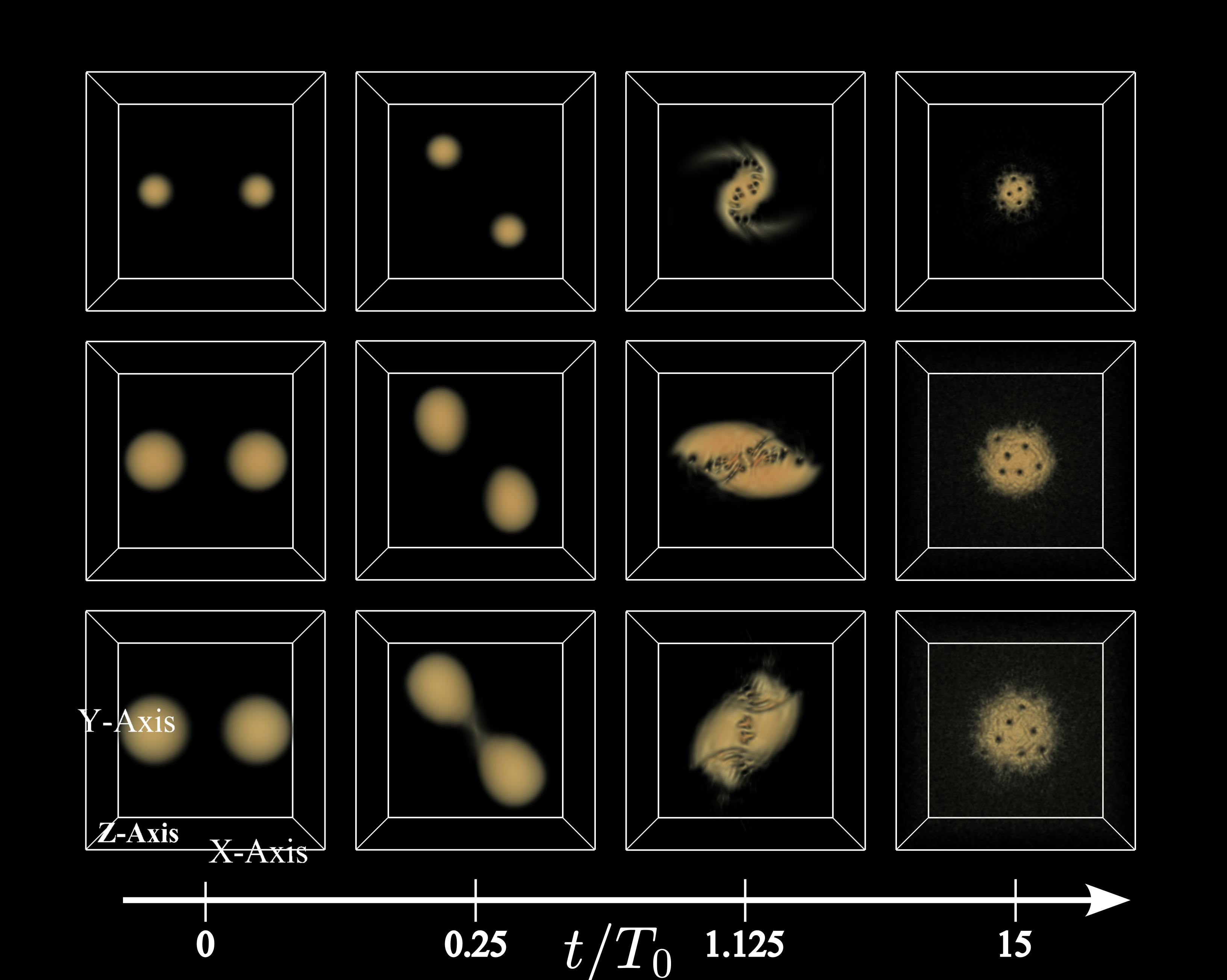}
   \caption{Pseudocolor density volume renderings of $|\psi|^2$, at representative dimensionless times (increasing from left to right), for binary mergers of boson (top panel), axion (middle panel), and q-boson (bottom panel) stars. The temporal scale $T_0$ is given in Appendix~\ref{app:model}. In the late stages of the mergers, vortical structures thread the merging objects.}
   \label{fig:image_binary_merger}
\end{figure*}



\parheading{Gravitational Waves:} Our binary mergers have weak internal gravity and slow internal motion; therefore, we compute the gravitational waves (GW) emission in the linearized, slow-motion Newtonian-quadrupole limit~\cite{finn1990determining, poisson2014gravity}. This approximation is well justified for our fiducial QCD-axion-scale ($M_{\star} = 10^{-12}M_\odot$, $R_{\star} \sim 10^3 $ km) giving the corresponding virial velocity $v/c= \sqrt{GM_{\star}/(R_{\star}c^2)}\simeq 10^{-8}$; a similar calculation for the ultralight axion halo yields $v/c \simeq 10^{-4}$[see Appendix~\ref{app:model}]. In this limit, the space-time metric can be linearized as $g_{\mu\nu} \approx \eta_{\mu\nu} + h_{\mu \nu}$~\cite{finn1990determining, poisson2014gravity}, where $\eta_{\mu\nu}$ is the flat space-time metric and $ h_{\mu \nu}$ is a small perturbation. The gravitational radiation is characterized by the transverse-traceless part of $h_{\mu \nu}$, the traceless quadrupole tensor $\mathcal{I}_{ij}$, and the gravitational power or luminosity $ \mathcal{L}$. In the slow-motion approximation these are defined as follows~\cite{finn1990determining, poisson2014gravity}:
\begin{eqnarray}
    h_{ij}^{TT} &=& \frac{G}{c^4}\frac{2}{D}\frac{d^2\mathcal{I}_{ij}}{dt^2}\,;\\
\mathcal{I}_{ij} &=& \int m|\psi|^2\left(r_ir_j-\tfrac{1}{3}\delta_{ij}r^2\right)d^3r\,;\\
\mathcal{L} &=& (G/5c^{5})\sum_{ij}\left| \dddot{\mathcal{I}}_{ij} \right|^2\,;
\end{eqnarray}
here, $D$ is the distance of the source from the observer.

We first define the complex strain 
\begin{eqnarray}
h(t) &=& h_+ + i h_{\times}\,; \nonumber \\
h_+ &\equiv& h_{xx} = -h_{yy} \quad {\rm and} \quad h_\times \equiv h_{xy}\,;
\label{eq:ht}
\end{eqnarray}
the subscripts $+$ and $\times$ denote plus and cross polarizations~\cite{blanchet2019analyzing}, respectively. 
We then measure the waveform along the orbital-angular-momentum axis; in the standard decomposition for quasi-circular non-precessing binaries~\cite{blanchet2019analyzing}, the leading contribution is the quadrupolar $(\ell,m)=(2,2)$ mode, and the GW strain is
\begin{equation}
    h_+(t)=A(t)\cos({\phi^{GW}})\,, \qquad h_{\times}(t)=A(t)\sin({\phi^{GW}})\,.
    \label{eq:phiGW}
\end{equation}
Using the phase of the equation $h = h_+ + i h_{\times}$, the instantaneous GW frequency is $f_{GW}=(1/2\pi)|\operatorname{Im}(\dot h/h)|$. The continuous wavelet transform (CWT) of $h$ identifies the dominant merger frequency $f_2$~\cite{bauswein2012measuring}, with the saturated scalogram maxima marking the inspiral-to-ringdown transition.




\parheading{Binary mergers:} We first present pictorial views of the mergers we consider; this is followed by the details of our analysis of the GW signatures from axion, q-boson, and boson compact objects. 
\begin{figure*}[htb!]
   \centering
   \includegraphics[width=2.0\columnwidth]{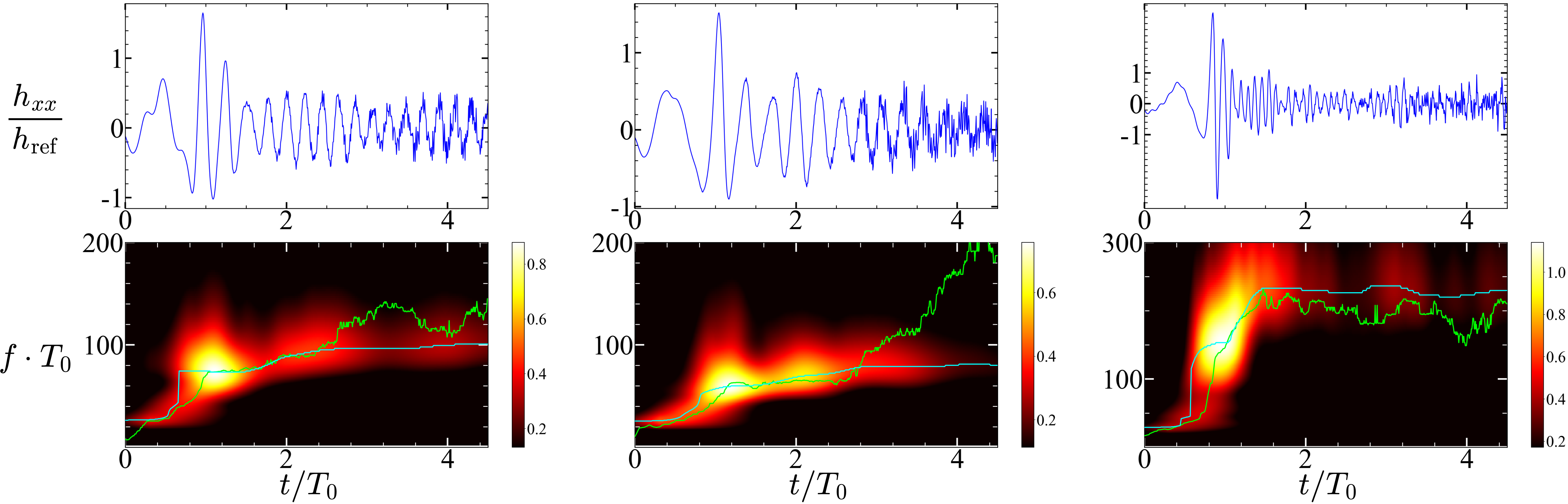}
   \caption{Top panel: Plots versus the dimensionless time $t/T_0$ of the scaled GW strain $h_{xx}/h_{ref}$ [see Eq.~\eqref{eq:ht}]; $h_{ref}=(1/D)\cdot 1.512\times10^{-169}M_0^{-4}L_0^{-3}$ (see text).   
   Bottom panel: Pseudocolor plots of the continuous wavelet transforms (CWTs) of $h(t)$ [see Eq.~\eqref{eq:ht}]; the cyan curves show the ridge frequency (i.e.,
   the peak of the scalogram at each instant of time); and the green curves depict the instantaneous GW frequency $f_{GW}=|d\phi^{GW}/dt|$ [see Eq.~\eqref{eq:phiGW}]. All plots are for representative values of $\Lambda$. Left panel: axion system ($\Lambda=-4.964$); middle panel: q-boson system ($\Lambda=4.964$); right panel: boson system  ($\Lambda = \infty$).}
   \label{fig:plots_binary_merger}
\end{figure*}
We simulate mergers starting from the initial condition~\eqref{eq:double_wavefunc}, with two identical stars in the binary system, for a spectrum of values of $\Lambda$ that interpolate from axion, through q-boson, to boson objects. We present results for three representative values of $\Lambda$: The top, middle, and bottom panels of Fig.~\ref{fig:image_binary_merger} show, respectively, pseudocolor volume renderings of $|\psi|^2$ that illustrate the mergers for a boson binary ($\Lambda = \infty$), an axion binary ($\Lambda=-4.964$), and a q-boson binary ($\Lambda=4.964$); henceforth, we do not repeat these values of $\Lambda$. In the late stages of the mergers, vortical structures thread the merging objects. These vortices capture the initial orbital angular momentum~\cite{shukla2024gravity, siemonsen2023binary}; and it has been suggested, in the context of dark-matter halos, that the formation of such vortices may be linked to observed dips in galactic rotation curves~\cite{feng2026rotation, sofue2013rotation, zinner2011vortex}. Gravity dominates the merging dynamics of the objects, so the orbital velocity [$|\mathbf{v}_1|$ in Eq.~\eqref{eq:double_wavefunc}] required for a stable binary is largest for q-bosons and smallest for bosons, because the radii of the collapsed objects are largest for q-bosons and smallest for bosons; the radii of axionic assemblies lie between these two extremes, given the parameter range we explore.

Figure~\ref{fig:plots_binary_merger} (top panel) shows the $+$ polarization strain $h_+=h_{xx}$  [see Eq.~\eqref{eq:ht}] versus the scaled time $t/T_0$ for the three mergers in Fig.~\ref{fig:image_binary_merger}, normalized by $h_{ref}=(1/D)\cdot 1.512\times10^{-169}M_0^{-4}L_0^{-3}$ [see Appendix~\ref{app:grav_wave}]. The bottom panel of fig.~\ref{fig:plots_binary_merger} presents the continuous wavelet transform (CWT) of $h(t)$ [see Eq.~\eqref{eq:ht}], with the instantaneous frequency $f_{GW}$ (green) overlaid on the CWT ridge (cyan); this ridge is extracted from the peak of the scalogram. The characteristic rising ridge frequency, referred to as a chirp, occurs during the inspiral phase; it appears in all the examples we consider [see Appendix~\ref{app:freq_plots}], and in the three cases shown in fig.~\ref{fig:plots_binary_merger}, namely, axion, boson, and q-boson. After the chirp, we observe, for all these cases, that the green and cyan curves in Fig.~\ref{fig:plots_binary_merger} begin to saturate; then there is a decay of the CWT amplitude in the post-merger phase. We focus on the most prominent saturated peak denoted by $f_2$ (also known as the dominant merger frequency), as considered in GW studies of binary mergers of black holes and neutron stars~\cite{bauswein2012measuring, abbott2016observation}. We identify $f_2$ as the frequency with the highest amplitude in the CWT. We note that for boson merger the $f_2$ is not apparent as the CWT maximum streches for a large range of frequencies and does not coincide with saturation of the frequency curves, and hence we cannot extract their dominant merger frequency consistently. The vanishing of CWT amplitude denotes the end of the merger phase, and we cannot perform frequency analysis further as the instantaneous frequency (green curve) also deviates from the ridge frequency (cyan curve) in Fig.~\ref{fig:plots_binary_merger}. In the post-merger phase, the gravitational energy radiated by the merger remanent object is prominently by its oscillations and the coalescing of the two merging stars.\\


\begin{figure}[H]
    \centering
    \includegraphics[width=\columnwidth]{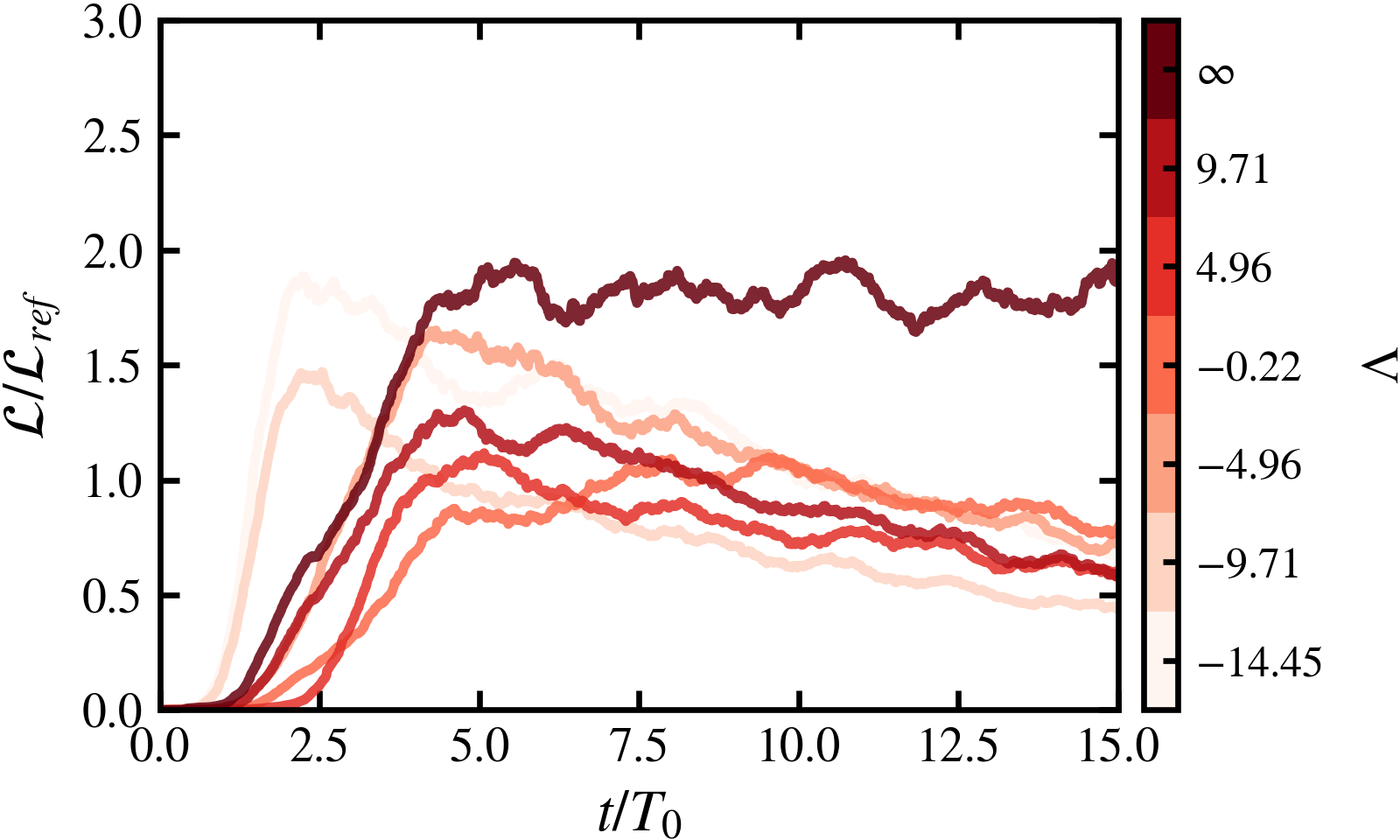}
    \caption{Plots versus the dimensionless time $t/T_0$ of the scaled GW luminosity $\mathcal{L}/\mathcal{L}_{ref}$; $\mathcal{L}_{ref} = 5.143\times10^{-372}M_0^{-10}L_0^{-10}\, \text{watts}$ (see text). The colorbar represents different values of the dimensionless interaction strength $\Lambda$.}
    \label{fig:L_plot}
\end{figure}

\parheading{GW Luminosity:} Next we calculate the normalized GW luminosity $\mathcal{L}/\mathcal{L}_{ref}$, i.e., the normalized GW power emitted [with $\mathcal{L}_{ref} = 5.143\times10^{-372}M_0^{-10}L_0^{-10}\, \text{watts}$ (see Appendix~\ref{app:grav_wave})]. In Fig.~\ref{fig:L_plot} we plot $\mathcal{L}/\mathcal{L}_{ref}$ versus the scaled time $t/T_0$; for negative $\Lambda$ this increases sharply with time and then decays after a prominent peak; for $\Lambda > 0$ the increase in  $\mathcal{L}/\mathcal{L}_{ref}$ and its peak are less
prominent; and for $\Lambda = \infty$, no decay is visible on the time scale of our DNS. The long-time behavior of $\mathcal{L}/\mathcal{L}_{ref}$ can be associated with the GW signal from the merger remanent. The time $t_{max}/T_0$ at which the main peak occurs in $\mathcal{L}/\mathcal{L}_{ref}$ depends on $\Lambda$; furthermore, $t_{max}/T_0$ is greater than the time at which the merger begins (column 3 of Fig.~\ref{fig:image_binary_merger}) but less than the time of complete coalescence  (column 4 of Fig.~\ref{fig:image_binary_merger}). We hypothesize that the long-time value of
$\mathcal{L}/\mathcal{L}_{ref}$ is large if the number of vortices in the merger remanent is small (as in the top boson row of Fig.~\ref{fig:image_binary_merger}).


\begin{figure}[H]
    \centering
    \includegraphics[width=\columnwidth]{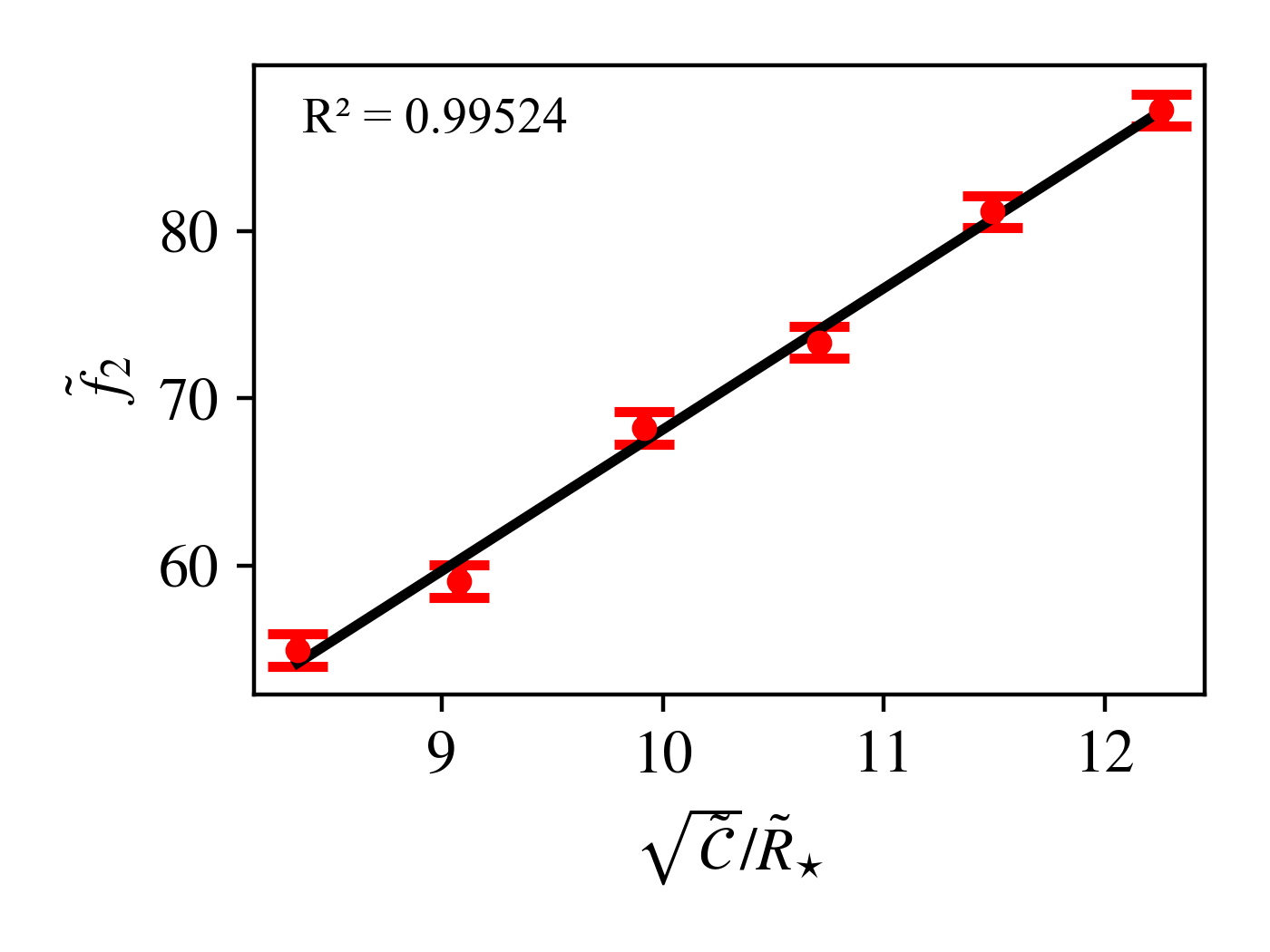}
    \caption{Dimensionless merger frequencies $\tilde f_2=T_0f_2$ plotted for each particle. Black line shows least-squares linear fit with $R^2=$ determination coefficient. Error bars represent the residual standard deviation. Inset figure includes the data point for boson in blue.}
    \label{fig:fvsC}
\end{figure}

\parheading{Compactness-Frequency analysis:} While the constituents of our binary system are spiraling towards each other, we can use the analysis that has been developed for neutron-star, and boson-star mergers to calculate the frequency of the inspiral [calculated as the frequency of the innermost stable circular orbit (subscript ISCO)~\cite{giudice2016hunting}]:
\begin{equation}
    f_{\rm ISCO}
    \propto \sqrt{\frac{M_\star}{R_\star^3}}
    \propto \frac{\mathcal{C}^{1/2}}{R_\star}\,,
    \label{eq:contact_freq}
\end{equation}
where $\mathcal{C}\equiv M_{\star}/(R_{\star})$ is defined as the \textit{compactness} of the star. It is natural to hypothesize that the dominant merger frequency $f_2$ [see Fig.~\ref{fig:plots_binary_merger}] is related to $f_{\rm ISCO}$; indeed, this hypothesis has been verified in a neutron-star study~\cite{takami2015constraining}. We show that this hypothesis holds also for the mergers we consider. 
In particular, we make the Ansatz

\begin{equation}
    f_{2} \sim f_{\rm ISCO}
    \propto \frac{\mathcal{C}^{1/2}}{R_\star}\,,
    \label{eq:contact_freq}
\end{equation}
and then test whether it is consistent with the results of our DNS.
This relation~\eqref{eq:contact_freq} tells that the dominant merger frequency depends on the merging objects only through $M_{\star}$ and $R_{\star}$. We verify Eq.~\eqref{eq:contact_freq} by keeping $\tilde{G}$ and $\tilde{g}_2$ fixed and varying only one parameter, namely, $\tilde{g}$ in $\Lambda= \tilde{g}/\sqrt{\tilde{g}_2}$. The values we use span the entire range from strongly attractive to repulsive 
quartic self-interactions. For the radius of the star we use  
\begin{equation}
\tilde R_{\star} \equiv \tilde R_g=\sqrt{\int m|\psi|^2 rdr\, \bigg/ \,\int m|\psi|^2dr}\,,
\label{eq:Rg}
\end{equation}
where $\tilde R_g$ is the dimensionless radius of gyration in our DNSs. Furthermore, we employ
the enclosed dimensionless mass of the star $\tilde{M_{\star}}$ to calculate the compactness 
\begin{equation}
\tilde{\mathcal{C}}=\tilde M_\star/\tilde R_\star\,,
\label{eq:compact}
\end{equation}
where tildes denote dimensionless values. Table~\ref{tab:lambda_compactness_radius} gives the values of the compactness and radius of gyration for different values of $\Lambda$. Although these coefficients yield different self-interactions at the microscale, they affect the final merger frequency only via $\tilde{\mathcal{C}}$.

 For various values of $\Lambda$, we obtain the dominant frequency $f_2$ that is indicated by dotted lines on the CWT plots shown in the Figure in the Appendix~\ref{app:freq_plots}. 
 We then obtain the dimensionless frequency $\tilde{f}_2=T_0f_2$ and plot it versus the dimensionless $\tilde{\mathcal{C}}^{1/2}/\tilde{R}_*$ in Fig.~\ref{fig:fvsC}; the red circles with error bars show data points from our DNS; the black line, a linear fit to these points, provides 
 a compelling verification of the relation~\eqref{eq:contact_freq}.
 Similar microphysics-independent frequency-compactness correlations have been reported for neutron-star mergers across different nuclear equations of state~\cite{takami2015constraining}. Our results show, for the first time, that the same relation~\eqref{eq:contact_freq} 
 holds for the non-relativistic scalar-field compact objects we consider.


\parheading{Analytical results via a Gaussian Ansatz:} In our DNSs, the merger frequency shows a dependence only on the compactness.
It is natural to ask whether the compactness itself, as a function of self-interaction coefficient $\Lambda$, can be obtained analytically rather than only via DNSs. We show below that this is possible if we use the following Gaussian Ansatz [superscript $\mathcal{A}$] for the wave function of the compact objects:
\begin{equation}
    \tilde\psi^\mathcal{A}(\tilde{r}) \equiv \frac{\sqrt{N}}{\pi^{3/4} \sigma^{3/2}}e^{-\tilde{r}^2/(2\sigma^2)}\,.
    \label{eq:anstaz_wavef}
\end{equation}
After evaluating the different terms in the total energy $E = E_{kin}+E_g+E_{g_2}+E_G$ [see Appendix~\ref{app:gaussian ansatz}], we get, in this Gaussian
approximation,
\begin{equation}
    E^\mathcal{A}(\sigma) = \frac{3N}{4\sigma^2} -\frac{\tilde G N^2} {4\sqrt{2}\pi^{3/2}\,\sigma} +\frac{\tilde gN^2} {4\sqrt{2}\,\pi^{3/2}\sigma^3} +\frac{\tilde g_2N^3} {9\sqrt{3}\,\pi^3\sigma^6}\,,
    \end{equation}
The mininmization of $E^\mathcal{A}(\sigma)$ yields the equilibrium condition
\begin{equation}
    -\frac{3N}{2\sigma^3} + \frac{\tilde GN^2} {4\sqrt{2}\pi^{3/2}\,\sigma^2} - \frac{3\Lambda\sqrt{\tilde g_2}\,N^2} {4\sqrt{2}\,\pi^{3/2}\sigma^4}- \frac{2\tilde g_2N^3} {3\sqrt{3}\,\pi^3\sigma^7}=0\,,
\label{eq:sigma_N}
\end{equation}
where $\tilde{g} = \Lambda \sqrt{\tilde{g}_2}$, and from which we obtain the following Gaussian-Ansatz results:
\begin{eqnarray}
\tilde{\mathcal{C}}^\mathcal{A} &\equiv& \tilde{M}^\mathcal{A}/\tilde{R}_g^\mathcal{A}\,;\quad
\tilde{R}_g^\mathcal{A} = \sqrt{3/2}\sigma\,; \nonumber\\
\tilde{M}^\mathcal{A} &\equiv & \int_0^{\tilde{R}_g^\mathcal{A}}\rho\ d^3 {\bf r}\,\Rightarrow \quad
\tilde{M}_g^\mathcal{A}\equiv f^\mathcal{A}N\,;\nonumber \\
f^\mathcal{A}&=& \mathrm{erf}(\sqrt{3/2})-\sqrt{6/\pi}e^{-3/2}\,.
\label{eq:GAres}
\end{eqnarray}

\begin{figure}
    \centering
    \includegraphics[scale=1]{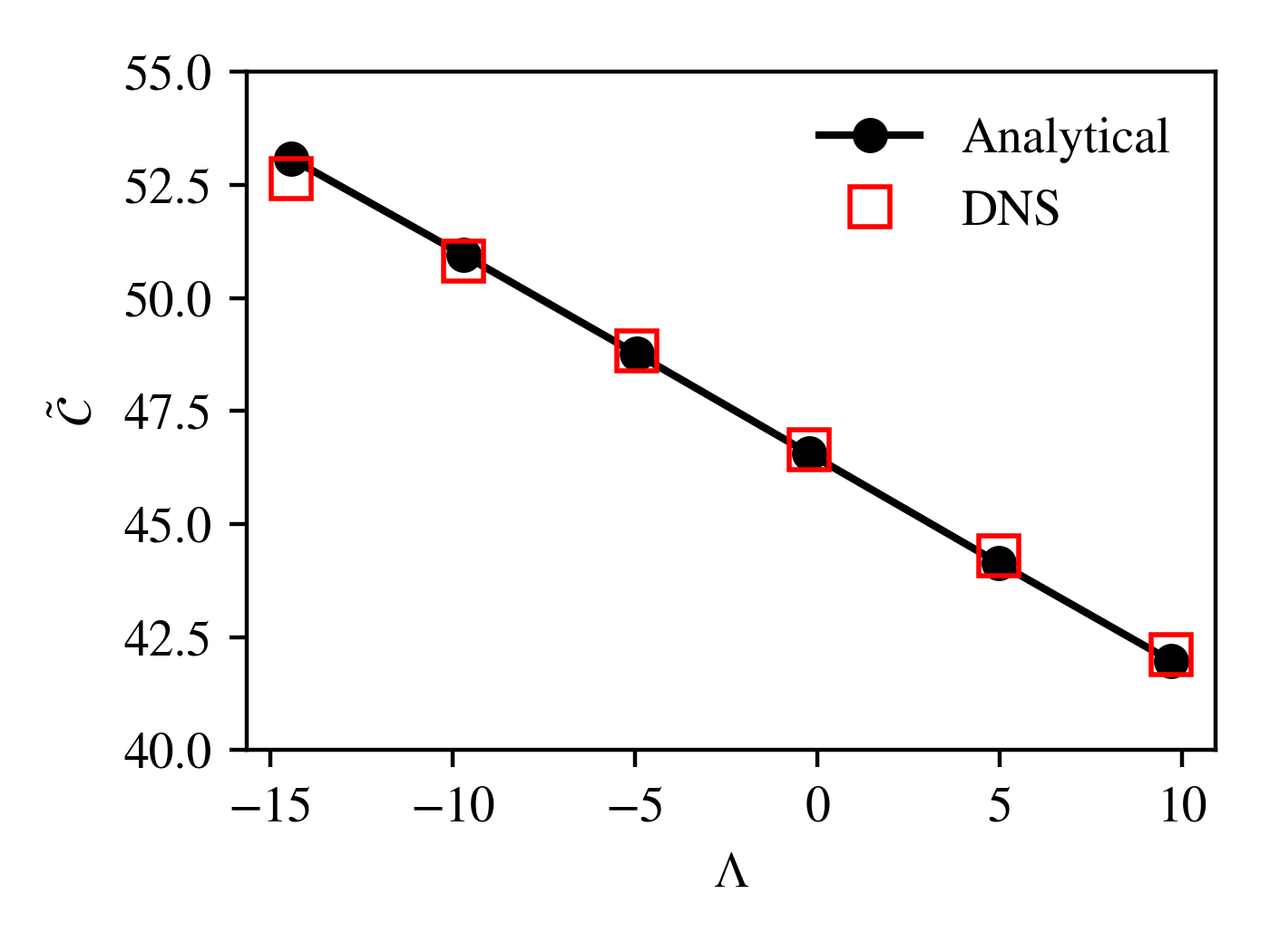}
    \caption{Dimensionless compactness $\tilde{\mathcal{C}}$ versus the self-interaction parameter $\Lambda$ using our DNSs (red squares). The black line indicates the result of our Gaussian Ansatz (see text).}
    \label{fig:compactness_analytical}
\end{figure}
For a fixed value of $\tilde{g}_2$, $\Lambda$ controls the quartic interaction. We solve Eq.~\eqref{eq:sigma_N} numerically, for different values of $\Lambda$, to obtain the equilibrium value of $\sigma$ and then, via Eq.~\eqref{eq:GAres}, the dimensionless radius of gyration $\tilde{R}_g^\mathcal{A}$ and the 
compactness $\tilde{\mathcal{C}}^\mathcal{A}$. The solid black line in
Fig.~\ref{fig:compactness_analytical} is a plot of $\tilde{\mathcal{C}}^\mathcal{A}$ versus $\Lambda$; our DNS results for $\tilde{\mathcal{C}}$,
indicated by red squares, are in good agreement with $\tilde{\mathcal{C}}^\mathcal{A}$. This shows that the merger frequency coming from Kepler's law depends on compactness, whose $\Lambda$-dependence itself arises from the elementary energy balance where gravity, quantum pressure, and self-interaction compete for the equilibrium radius.
 


\parheading{Discussion and Conclusions:} We have studied equal-mass binary mergers of self-gravitating scalar-field compact objects across a spectrum of attractive and repulsive self-interaction within the cubic-quintic Gross-Pitaevskii-Poisson framework. Using a dimensionless interaction parameter $\Lambda$, we obtain a unified description connecting axion-like ($\Lambda<0$) and boson-like ($\Lambda>0$) systems. We have demonstrated via our DNSs, that the gravitational-wave (GW) emission and post-merger dynamics exhibit a systematic dependence on the nature and strength of the self-interaction.

We first show that the evolution of the GW luminosity is sensitive to the nature of the self-interaction. Mergers with attractive quartic self-interactions exhibit a pronounced decay in their post-merger GW luminosity, whereas repulsive configurations tend towards sustained emission from the merger remanent. This behavior highlights the role of microscopic interactions in shaping the internal dynamics and subsequent evolution of the merger. In contrast, the dominant merger frequency $f_2$ depends only on the compactness of the merging objects. Across the range of interaction parameters that we have studied, the dimensionless merger frequency follows the scaling law $\tilde{f}_2\sim(\widetilde{C}^{1/2}/\widetilde{R}_{\star})$, which is consistent with the expected Keplerian dependence of orbital frequency on mass and radius. Thus, although the self-interactions govern the equilibrium structure and the evolution of the GW luminosity, $\tilde{f}_2$ follows from a single unified macroscopic parameter.

To understand the origin of this unified relation, we have developed a Gaussian Ansatz and used it to obtain an analytical relation between its compactness and the self-interaction parameter. The results of this Ansatz are in good agreement with their counterparts from our DNS. This agreement explains how gravity, quantum pressure, and self-interaction collectively determine the characteristic size and compactness of the merger remanent.

The physical scales for our fiducial axion-motivated configurations correspond to an extremely small compactness and weak internal gravitational fields. Consequently, the associated GW signals are not expected to be detectable by current ground-based detectors. This limitation should be distinguished from the existence of potentially interesting waveform signatures, where the simulation, using dimensionless quantities, reveals systematic differences in the merger dynamics across the spectrum of self-interaction values. In particular, the compactness-frequency relation, in conjunction with the analytical compactness relation obtained using a Gaussian Ansatz, offers a possible route toward the interpretation of merger frequencies in terms of the underlying scalar-field interactions. The results of our study should be viewed primarily as theoretical predictions for the GW signatures of weak-field 
scalar-field compact-object mergers rather than as predictions for currently detectable events. Our work provides a strong motivation for future 
efforts to constrain dark-matter interactions, which must eventually employ a relativistic treatment for boson star-black hole binaries.




\parheading{Acknowledgments:}  We thank the Anusandhan National Research Foundation (ANRF), the Science and Engineering Research Board (SERB), for support,  and the Supercomputer Education and Research Centre (IISc), for computational resources.

\bibliographystyle{unsrt}

\appendix

\section{Model and methods}
\label{app:model}
Our simulation domain is cubical box, with a side of (non-dimensional) length $L=2\pi$~\cite{canuto2012spectral,shukla2024gravity}, for which we use periodic boundary conditions in all three directions. 
For this DNS, we use Fourier expansions of the fields: $\Psi \equiv (\psi,\Phi); \quad \Psi ({\bf x}) = \sum_{ {\bf k} } \hat{\Psi}_{{\bf k}} \exp (i {\bf k} \cdot {\bf x})$. It is natural, therefore, to evaluate derivatives in Fourier space;
but, to bypass computationally costly convolutions, we obtain products in physical space. To suppress aliasing errors arising from the evaluation of nonlinear terms in the pseudospectral method, we employ the standard $2/3$ dealiasing rule, $\hat{\Psi}_{\bf k}=0$ for $|{\bf k}| > N/3$~\cite{OntheEliminationofAliasinginFiniteDifferenceSchemesbyFilteringHighWavenumberComponents,canuto2012spectral}. For time evolution we employ the fourth-order Runge-Kutta (RK-4) method. We use $N=256^3$ collocation points; to 
show that our results are not limited significantly by our spatial resolution, we compare our results with those from a DNS with $128^3$ collocation points [see Appendix~\ref{app:N=128}].

We prepare our initial condition by obtaining ground-state solutions, consisting of a single star, using the imaginary-time relaxation ($t\to-it$) for Eqs.~\eqref{eq:GPP_equation} which is the following Stochastic Ginzburg–Landau–Poisson Equation (SGLPE):
\begin{eqnarray}
\hbar \frac{\partial }{\partial t}\psi(\mathbf{r}, t) &=& \bigg( \frac{\hbar^2}{2m} \nabla^2 - g |\psi(\mathbf{r}, t)|^2 - g_2 |\psi(\mathbf{r}, t)|^4 \nonumber\\
&&- m\Phi \bigg) \psi(\mathbf{r}, t)\,;
\label{eq:imag_GPP}
\end{eqnarray}
for this SGLPE~\eqref{eq:imag_GPP}, we use the initial condition that has a uniform density distribution for $\psi$, with the following small-amplitude perturbation at low wavenumbers $k$:
\begin{equation}
    \psi({\bf r},t=0)= \begin{cases}
                        1, & k=0 \,;\\
                        \frac{10^{-2}}{k^2}, & 1<|k|<10\,;\\
                        0, & 10\leq|k|\,.
                       \end{cases}\,
\label{eq:initial_condition}
\end{equation}
The resulting single-star density profiles and their fit against the empirical Schive profile~\cite{schive2014cosmic} are given in the Supplemental Material.

To initialize our binary systems, we place the centers of two such identical single-star density profiles on the $x$-axis at $\pm {\bf x_0}$, with symmetrical orbital velocities (${\bf v}_2\ =-{\bf v}_1$) in the $XY$ plane, so the total wavefunction of the system becomes:
\begin{equation}
\psi_b(\mathbf{x},t=0) = f(|\mathbf{x}-\mathbf{x}_0|)e^{i\mathbf{v}_1\cdot\mathbf{x}} + f(|\mathbf{x}+\mathbf{x}_0|)e^{-i\mathbf{v}_1\cdot\mathbf{x}}\,.
\label{eq:double_wavefunc}
\end{equation}
We can introduce a relative phase between the two terms on the right-hand side of Eq.~\eqref{eq:double_wavefunc}; we choose the relative phase to be zero because this leads to fast mergers~\cite{palenzuela2007head}.

We solve the the GPP equations~\eqref{eq:GPP_equation} by non-dimensionalizing them using the particle mass scale $m=M_0$ and a natural time scale $T_0= mL_0^2/\hbar$; here, $M_0$, $L_0$, and $T_0$ are the mass, length, and time scales, respectively. The resulting non-dimensional equation is [see Appendix~\ref{app:Dimensionless GPPE}]
\begin{eqnarray}
i\frac{\partial\tilde\psi}{\partial\tilde t}&=&\left[-\frac12\tilde\nabla^2+\tilde g|\tilde\psi|^2+\tilde g_2|\tilde\psi|^4
+\tilde\Phi\right]\tilde\psi\,;\nonumber\\
\tilde\nabla^2\tilde\Phi&=&\tilde Gm|\tilde\psi|^2\,.
\label{eq:gpe_nondim}
\end{eqnarray}
If we choose a target mass $M_\star$ for the star, then the length scale is taken to be the soliton radius $R_\star$, determined by the weakly interacting (dilute) axion-star mass--radius relation~\cite{chang2024axion, chavanis2011mass}:
\begin{equation}
R_\star \approx 2.6\times10^5\,\mathrm{km}
\left(\frac{10^{-6}\,\mathrm{eV/c^2}}{m}\right)^2
\left(\frac{10^{-12}M_\odot}{M_\star}\right);
\label{eq:radius_star}
\end{equation}
here, $M_\odot$ denotes the solar mass. For a QCD axion if we choose $m=10^{-5}\,\mathrm{eV/c^2}$ and, in addition,  we choose $M_\star=10^{-12}M_\odot$, then $R_\star\simeq 2.6\times10^3$ Km. This gives the corresponding virial velocity $v/c= \sqrt{GM_{\star}/(R_{\star}c^2)}= 2.38\times 10^{-8}$. By contrast, for ultralight axions with particle mass $m\simeq 10^{-24}\mathrm{eV/c^2}$, if we consider an ultralight-axion halo $M_{\text{halo}}\simeq 10^{11}M_\odot, \ R_{\text{halo}}\simeq 84.25 \ \mathrm{kpc}$ and the virial velocity is $v/c= \sqrt{GM_{\text{halo}}/(R_{\text{halo}}c^2)}= 2.38\times 10^{-4}$[for non self-interacting ultralight axions see Ref.~\cite{Hui2017ultralight}]. Note that the dimensionless gravitational constant used in our DNS equations is $\tilde G =4\pi Gm^2M_\star R_\star/\hbar^2$ in Eq.~\eqref{eq:radius_star}. 

\section{Jeans Instability and Conserved Quantities}
\label{app:GPPE}
If we linearize Eq.~\eqref{eq:GPP_equation} about the uniform state $|\psi|^2=n_0$, we get the dispersion relation
\begin{eqnarray}
    \omega(k) = \sqrt{\frac{\hbar^2k^4}{(2m)^2} -\frac{k^2}{m} \left(\frac{4\pi G m^2n_0}{k^2} -gn_0 -2g_2n_0^2\right)}\,
\label{eq:disprel}
\end{eqnarray}
where $k$ is the magnitude of the wavenumber. Using Eq.~\eqref{eq:disprel}, we obtain the Jeans wavenumber
\begin{equation}
\begin{aligned}
    k_J^2 &=& \frac{2m(gn_0+2g_2n_0^2)}{\hbar^2} \bigg[-1+ \sqrt{1+ \frac{4\pi G\hbar^2 m^2n_0}{(gn_0+2g_2n_0^2)^2}} 
 \bigg]\,;
 \end{aligned}
\end{equation}
if the wavenumber is less than $k_J$, the Jeans instability sets in; it then initiates the collapse of the mass distribution~\cite{harko2019jeans,chen2021new}.

Equations~\eqref{eq:GPP_equation} conserve the number of particles $N$ and the total energy $E$ which are, respectively:
\begin{eqnarray}
    N &=& \int|\psi|^2 d^3{\bf r}\,;\quad E = E_k + E_{SI} + E_G\,;\nonumber\\
     E_k &=& \int \frac{\hbar^2}{2m} |\nabla \psi|^2 d^3 {\bf r}\,;\nonumber\\
    E_{SI} &=& \int \left(\frac{g}{2}|\psi|^4+\frac{g_2}{3}|\psi|^6\right)d^3{\bf r}\,;\nonumber\\
    E_G &=& \frac{4\pi Gm^2}{2}\int  |\psi|^2 \nabla^{-2} |\psi|^2 d^3{\bf r}\,.
    \label{eq:N&E}
\end{eqnarray}
Here, $E_k$, $E_{SI}$, and $E_G$ are the kinetic, self-interaction, and gravitational energies, respectively.

\section{DNS Results for $N=128^3$}
\label{app:N=128}
\begin{figure}[H]
    \centering
    \includegraphics[width=\columnwidth]{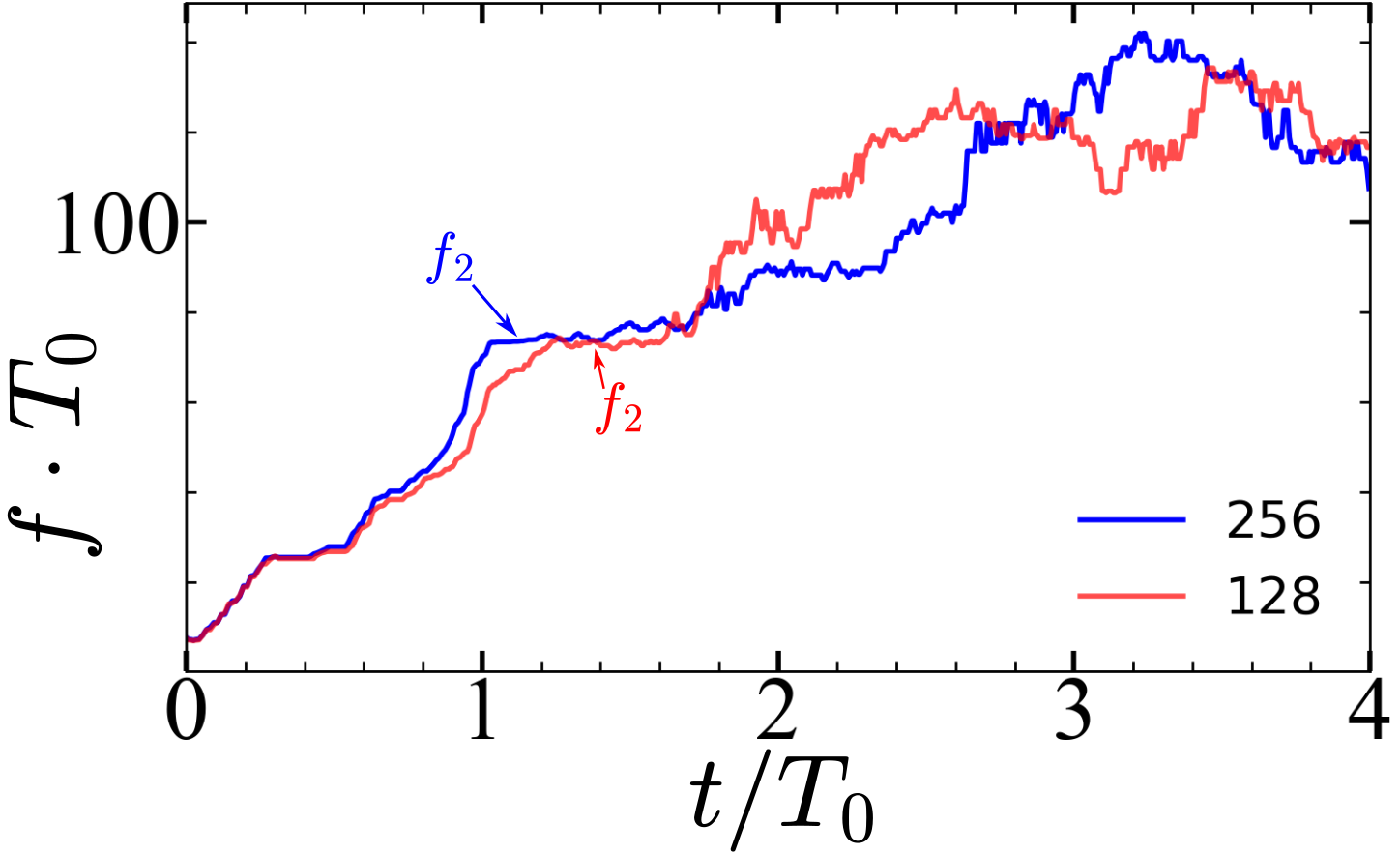}
    \caption{Instantaneous GW frequency plots for axion-star mergers from our DNSs with $N=256^3$ and $N=128^3$.}
    \label{fig:f_comp}
\end{figure}

To validate the numerical fidelity of our code, we compare the frequency plots with $1/8$ the resolution of the DNSs presented in the main paper
(i.e. with  $N=128^3$). The instantaneous GW frequency $f_{GW}=(1/2\pi)|\operatorname{Im}(\dot h/h)|$, when plotted for binary star mergers from the same axion particle, show similar behaviors for both $N=128^3$ and $N=256^3$ DNSs. Using the same initial condition for simulations with $N=256^3$ and $N=128^3$ grid points, we show in Fig.~\ref{fig:f_comp} that the lower resolution simulation has the same dominant merger frequency $f_2$.

\section{Density profiles for the collapsed object}
\label{app:dens_profile}
\begin{figure*}[htb!]
   \centering
   \includegraphics[width=1.9\columnwidth]{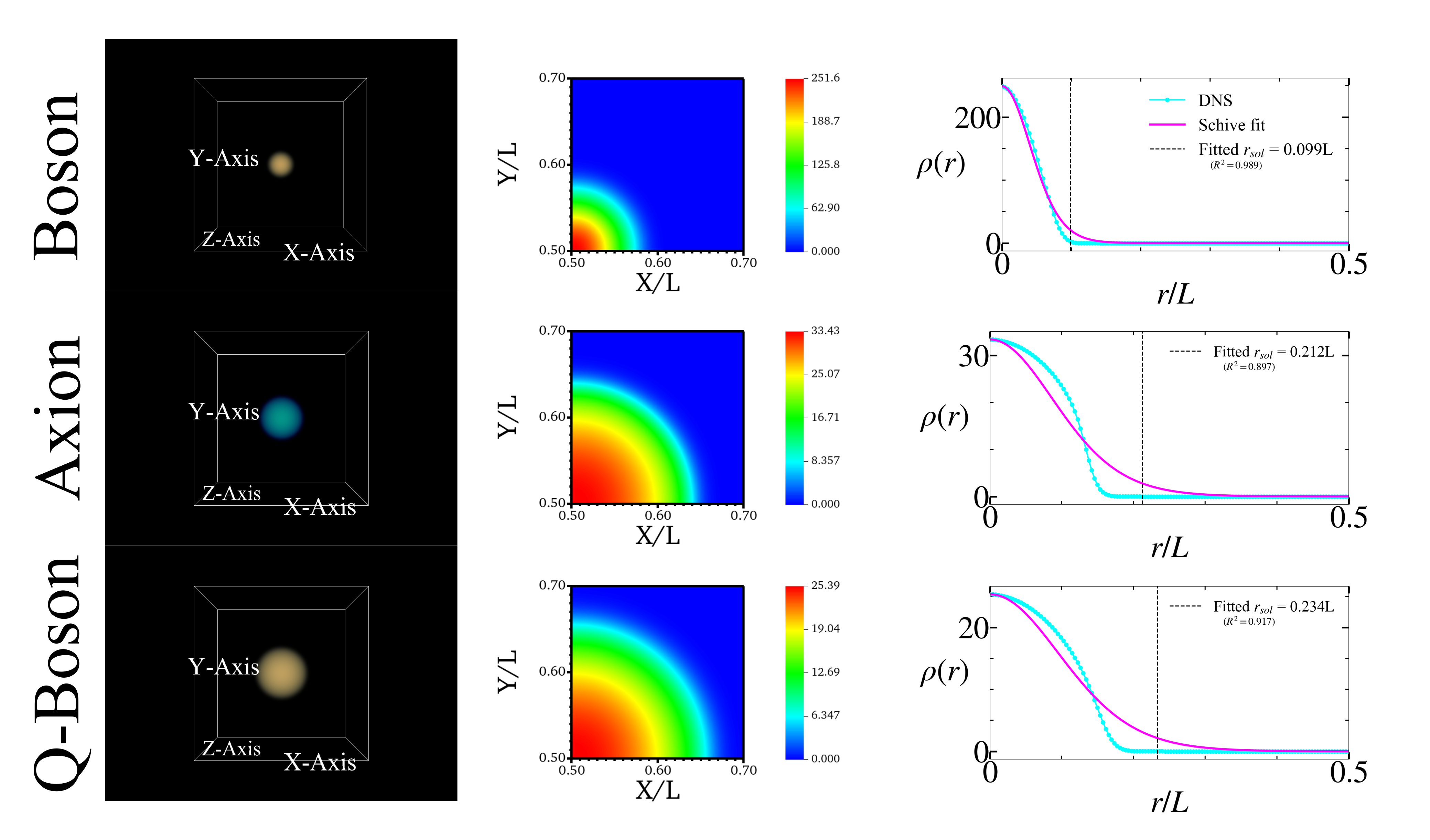}
   \caption{Left panel: One level contour plot of the density $|\psi|^2$. Middle panel: Pseudocolor density plot at the plane $z=L/2$. Right panel: spherically averaged density $\rho$ as a function of the distance from the center $r$ divided by box length $L$ fitted against \textit{Schive} profile }
   \label{fig:single_axion_star}
\end{figure*}
We obtain a single spherically collapsed object (single star) for each of the three values of $\Lambda$ in the paper. In Fig.~\ref{fig:single_axion_star}
(left panel) we show the single level contour plot of the BEC density $|\psi|^2$. In Fig.~\ref{fig:single_axion_star} (middle panel) we show the cross section for each of the star in the $XY$ plane. The presence of the sextic repulsion term, in axion and q-boson interactions, leads to a comparatively larger radius of the collapsed object (star) than for its boson counterpart.
The core density of our boson star, formed as a condensate for the  cubic GPPE, can be fitted with an empirical Schive profile~\cite{Schive2014}; this is given in Eq.~\eqref{eq:schive profile}, where $r_{sol}$ is the width of the density profile and $\rho_c$ is the density at the star-center. In Fig.~\ref{fig:single_axion_star} (right panel) we plot the star's spherically averaged radial density $\rho(r) = m|\psi|^2$, shown in cyan; 
the magenta curve shows the best fit to the Schive-profile.
\begin{eqnarray}
    \rho_{\small\text{fit}}(r) = \frac{\rho_c}{\left[1+0.091(\frac{r}{r_{sol}})^2\right]^8}\,.
    \label{eq:schive profile}
\end{eqnarray}

\section{Dimensionless GPPE}
\label{app:Dimensionless GPPE}

To convert Eq.~\eqref{eq:GPP_equation} in a non-dimensionalized form used in our code we introduce the length scale $L_0$, the time scale $T_0$, and the wavefunction amplitude $\Psi_0$. Thus, our dimensionless simulation scales $(\tilde{\mathbf r}, \tilde t, \tilde\psi)$ can be converted to physical scales with the relations given below:
\begin{eqnarray}
   \mathbf{r} &=& L_0\tilde{\mathbf r},\nonumber\\
t&=&T_0\tilde t,\nonumber\\
\psi(\mathbf r,t)&=&\Psi_0\tilde\psi(\tilde{\mathbf r},\tilde t).
\label{eq:scalings} 
\end{eqnarray}
The differential operators thus transform as
\begin{eqnarray}
    \nabla=\frac{1}{L_0}\tilde\nabla,
\qquad
\nabla^2=\frac{1}{L_0^2}\tilde\nabla^2,
\qquad
\frac{\partial}{\partial t} =\frac{1}{T_0}\frac{\partial}{\partial\tilde t}.
\label{eq:non-dim_derivatives}
\end{eqnarray}

Substituting these relations into Eq.~\eqref{eq:GPP_equation} gives
\begin{align}
i\hbar\frac{\Psi_0}{T_0}
\frac{\partial\tilde\psi}{\partial\tilde t}=
\bigg[-\frac{\hbar^2}{2mL_0^2}\tilde\nabla^2 +g\Psi_0^2|\tilde\psi|^2 +g_2\Psi_0^4|\tilde\psi|^4\nonumber\\
+m\Phi_0\tilde\Phi \bigg]\Psi_0\tilde\psi.
\label{eq:substituted}
\end{align}
The natural kinetic energy scale associated with the length scale $L_0$ is
\begin{equation}
E_0=\frac{\hbar^2}{mL_0^2}\,;
\label{eq:energy-scale}
\end{equation}
and the corresponding natural time scale is
\begin{equation}
T_0=\frac{\hbar}{E_0}=\frac{mL_0^2}{\hbar}.
\label{eq:time-scale}
\end{equation}

Now, we choose $\Phi_0$ and $\tilde\Phi$ such that
\begin{eqnarray}
    \Phi_0 = \frac{\hbar^2}{m^2L_0^2}\,; \qquad \Phi = \Phi_0 \tilde\Phi\,.
    \label{eq:phi_not}
\end{eqnarray}
Substituting these relations in Eq.~\eqref{eq:substituted} and then dividing it by $\frac{\hbar^2\Psi_0}{mL_0^2 }$, we obtain
\begin{equation}
i\frac{\partial\tilde\psi}{\partial\tilde t}=\left[-\frac12\tilde\nabla^2+\tilde g|\tilde\psi|^2+\tilde g_2|\tilde\psi|^4
+\tilde\Phi\right]\tilde\psi.
\label{eq:dimensionless-gppe}
\end{equation}

Now, starting from the Poisson Eq.~\eqref{eq:GPP_equation},
\begin{equation}
\tilde\nabla^2\tilde\Phi
=
\tilde G
\left(
|\tilde\psi|^2-
\langle|\tilde\psi|^2\rangle
\right)\,,
\end{equation}
and using Eqs.~\eqref{eq:scalings} and \eqref{eq:phi_not}, we obtain
\begin{equation}
\tilde\nabla^2\tilde\Phi=\frac{4\pi Gm^3L_0^4\Psi_0^2}{\hbar^2}
|\tilde\psi|^2.
\end{equation}
Thus,
\begin{equation}
\tilde\nabla^2\tilde\Phi=
\tilde G|\tilde\psi|^2\,.
\label{eq:dimensionless-poisson}
\end{equation}
Equations~\eqref{eq:dimensionless-gppe} and \eqref{eq:dimensionless-poisson} constitute the dimensionless Gross-Pitaevskii-Poisson system. The dimensionless interaction coefficients in the GPPE are, therefore,
\begin{eqnarray}
    \tilde{g} &=& \frac{m g \Psi_0^2 L_0^2}{\hbar^2}\,;
    \qquad
    \tilde{g}_2 = \frac{m g_2 \Psi_0^4 L_0^2}{\hbar^2}\,;
    \nonumber \\
    \tilde{G} &=& \frac{4\pi G m^3 L_0^4 \Psi_0^2}{\hbar^2}\,.
    \label{eq:g2tilde-general}
\end{eqnarray}
Furthermore, in our simulations we normalize the dimensionless wavefunction via 
\begin{equation}
\int |\tilde\psi|^2d^3\tilde r=1.
\label{eq:dimensionless-normalization}
\end{equation}
Finally we get
\begin{align}
\ N&=\int \Psi_0^2|\tilde\psi|^2L_0^3d^3\tilde r\nonumber\\
\frac{M_*}{m}&=\Psi_0^2L_0^3\int |\tilde\psi|^2d^3\tilde r\nonumber\\
\Psi_0^2&=\frac{M_*}{mL_0^3}\,.
\label{eq:Psi0}
\end{align}

Substituting Eq.~\eqref{eq:Psi0} into the dimensionless interaction coefficients gives
\begin{equation}
\tilde g=\frac{gM_*}{\hbar^2L_0}, \qquad \tilde g_2=\frac{g_2M_*^2}{m\hbar^2L_0^4}.
\label{eq:g2tilde-normalized}
\end{equation}
Similarly, the dimensionless gravitational coupling becomes
\begin{equation}
\tilde G=\frac{4\pi Gm^2M_*L_0}{\hbar^2}.
\label{eq:Gtilde-normalized}
\end{equation}

The coefficients of $|\psi|^2$ and $|\psi|^4$ in Eq.~\eqref{eq:GPP_equation} are
\begin{eqnarray}
 g=\frac{4\pi a\hbar^2}{m^2}, \qquad g_2=\frac{32\pi^2 \hbar^4 a^2}{3 m^5 c^2} \,,
\end{eqnarray}
whence we get 
\begin{equation}
\frac{g^2}{g_2}=\frac{3m c^2}{2}.
\end{equation}
Using Eq.~\eqref{eq:g2tilde-normalized},
\begin{eqnarray}
    \frac{\tilde g^2}{\tilde g_2}=\frac{mL_0^2g^2}{g_2\hbar^2}
    =\frac{3}{2}\left(\frac{mcL_0}{\hbar}\right)^2 \equiv \Lambda^2.
\end{eqnarray}
The dimensionless speed of light in simulation scales is
\begin{equation}
\tilde c \equiv \frac{cT_0}{L_0}=\frac{mcL_0}{\hbar},
\label{eq:ctilde-def}
\end{equation}
where we used $T_0=mL_0^2/\hbar$. In terms of $\tilde c$,
\begin{equation}
\Lambda^2=\frac{3\tilde c^{\,2}}{2}
\qquad\Longleftrightarrow\qquad
\tilde c=\sqrt{\frac{2}{3}}|\Lambda| .
\end{equation}

\section{Gravitational Waves}
\label{app:grav_wave}
The gravitational radiation is completely given by the transverse-traceless part of $h_{\mu \nu}$; in the slow-motion approximation this
is 
\begin{equation}
    h_{ij}^{TT} = \frac{G}{c^4}\frac{2}{D}\frac{d^2\mathcal{I}_{ij}}{dt^2}\,,
\end{equation}
where $D$ is the distance of the source from the observer and $\mathcal{I}_{ij}$ is the traceless quadrupole tensor:
\begin{eqnarray}
    \mathcal{I}_{ij} &=& m\int |\psi|^2 (r_ir_j-\frac{1}{3}\delta_{ij}{\bf r}^2)d^3{\bf r}\,; \\
    \frac{d \mathcal{I}_{ij}}{dt} &=& 2m \int |\psi|^2 \  [ \frac{1}{2}(v_ir_j + v_jr_i)-\frac{1}{3}\delta_{ij}v_kr^k ] d^3{\bf r}\,.
    \label{eq:mass_quadrupole}
\end{eqnarray}
To obtain the measurable physical strain we restore the dimensions. The simulation scales are fixed by the boson mass $m\equiv1M_0$, chosen length scale $L_0$ and the natural time scale $T_0=\frac{M_0L_0^2}{\hbar}$. Using Eq.~\eqref{eq:mass_quadrupole} we can get quadrupole tensor in our scales as:
\begin{equation}
    \mathcal{I}_{ij}=M_*L_0^2\,\tilde{\mathcal{I}}_{ij}, \qquad
    M_*=\frac{\tilde G\,\hbar^2}{4\pi G\,M_0^2L_0},
\end{equation}
where the second relation is Eq.~\eqref{eq:Gtilde-normalized} solved for $M_*$. Then
\begin{equation}
    \frac{d^2 \mathcal{I}_{ij}}{dt^2}=\frac{M_*L_0^2}{T_0^2}\,\frac{d^2 \tilde{\mathcal{I}}_{ij}}{d\tilde t^2}
    =\frac{\tilde G\,\hbar^4}{4\pi G\,M_0^4L_0^3}\,\frac{d^2 \tilde{\mathcal{I}}_{ij}}{d\tilde t^2},
\end{equation}
and $G$ cancels in the strain, leaving
\begin{eqnarray}
    h_{xx}&=&\frac{\tilde G}{2\pi}\,\frac{\hbar^4}{c^4}\,\frac{1}{D}\,M_0^{-4}L_0^{-3}\,\frac{d^2\tilde{\mathcal{I}}_{xx}}{d\tilde t^2},\nonumber\\
    &=&\frac{1}{D}\,\bigl(1.512\times10^{-169}\bigr)\,M_0^{-4}L_0^{-3}\,\frac{d^2\tilde{\mathcal{I}}_{xx}}{d\tilde t^2}.
    \label{eq:Gravitational strain}
\end{eqnarray}
Here $M_0$ is in kg, and $L_0$ and $D$ are in meters.

The gravitational luminosity (power radiated) is calculated using the contribution of all the components of $\mathcal{I}_{ij}$:
\begin{eqnarray}
    \mathcal{L} &=& {\frac{G}{5c^{5}}}\sum_{ij}\left| \frac{d^3 \mathcal{I}_{ij}}{dt^3} \right|^2\nonumber\\
    &=& \frac{G}{5c^{5}}\,\frac{M_*^2L_0^4}{T_0^6}\sum_{ij}\left| \frac{d^3 \tilde{\mathcal{I}}_{ij}}{d\tilde t^3} \right|^2\nonumber\\
    &=& \frac{\tilde G^2\,\hbar^{10}}{80\pi^2\,G\,c^5}\,M_0^{-10}L_0^{-10}
    \sum_{ij}\left| \frac{d^3 \tilde{\mathcal{I}}_{ij}}{d\tilde t^3} \right|^2\nonumber
\end{eqnarray}
\begin{equation}
    \mathcal{L} = \bigl(5.143\times10^{-372}\bigr)\,M_0^{-10}L_0^{-10}
    \sum_{ij}\left| \frac{d^3 \tilde{\mathcal{I}}_{ij}}{d\tilde t^3} \right|^2
    \text{(watts)}.
    \label{eq:luminosity}
\end{equation}

\section{$\tilde{\mathcal{C}}$ and $\tilde R_\star$ for different values of $\Lambda$}
\begin{table}[htbp]
\centering
\caption{Dependence of compactness and radius on the
self-interaction parameter $\Lambda$.}
\label{tab:lambda_compactness_radius}
\begin{tabular}{ccc}
\hline
$\Lambda$ & $\tilde{C}$ & $\tilde{R_\star}/L\ \ (L=2\pi)$ \\
\hline
$-14.449$ & $52.64$ & $0.592$ \\
$-9.707$  & $50.82$ & $0.620$ \\
$-4.964$  & $48.83$ & $0.653$ \\
$-0.221$  & $46.65$ & $0.689$ \\
$+4.964$  & $44.30$ & $0.733$ \\
$+9.707$  & $42.12$ & $0.778$ \\
$\infty$  & $92.76$ & $0.370$ \\
\hline
\end{tabular}
\end{table}

\section{Frequency plots}
\label{app:freq_plots}
\begin{figure*}[htb!]
   \centering
   \includegraphics[width=1.9\columnwidth]{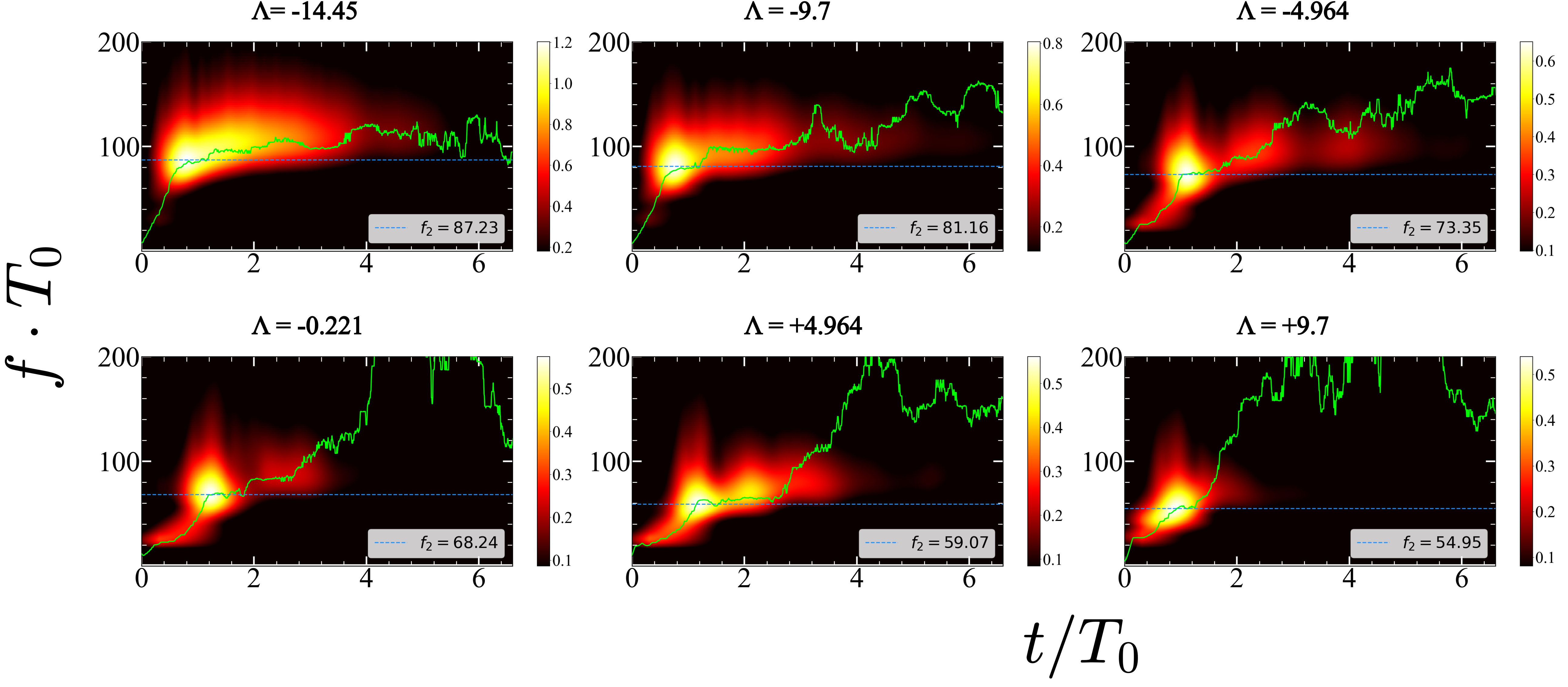}
   \caption{Pseudocolor plots of the continuous wavelet transforms (CWTs) of $h(t)$ [see Eq.~\eqref{eq:ht}]; the blue dotted lines show the scalogram maximum (i.e., the dominant merger frequency $f_2$); and the green curves depict the instantaneous GW frequency $f_{GW} = |d\phi^{GW} /dt|$ [see Eq. (10)]. The representative values of $\Lambda$ are shown on top of each plot.}
   \label{fig:frequency_analysis}
\end{figure*}

The characteristic rising ridge frequency, referred to as a chirp, appears in all the examples we consider. The CWT scalogram of $h(t)$ for different value of interaction coefficient $\Lambda$ is shown in Fig.~\ref{fig:frequency_analysis}. The scalogram peak denoted as dominant merger frequency $f_2$ is shown as a blue dotted line in plotted on top of the scalogram along with the green curve showing the instantaneous GW frequency $f_{GW} = |d\phi^{GW} /dt|$

\section{Gaussian Ansatz}
\label{app:gaussian ansatz}

To derive compactness from an analytical approach, we take our Ansatz wavefunction as:

\begin{equation}
\tilde\psi(\tilde{\mathbf{r}})
=
\frac{\sqrt{N}}{\pi^{3/4}\sigma^{3/2}}
e^{-\tilde{r}^2/(2\sigma^2)}.
\end{equation}

And our dimensionless GPP equation is:

\begin{equation}
i\frac{\partial\tilde\psi}{\partial\tilde t}
=
\left[
-\frac12\tilde{\nabla}^2
+\tilde g|\tilde\psi|^2
+\tilde g_2|\tilde\psi|^4
+\tilde\Phi
\right]\tilde\psi.
\end{equation}

with the respective dimensionless energy as:

\begin{equation}
E=E_{\rm kin}+E_g+E_{g_2}+E_G.
\end{equation}

Next, we calculate each energy separately, with kinetic energy $E_{\rm kin}$ as:

\begin{equation}
E_{\rm kin}
=
\langle\tilde\psi|\hat T|\tilde\psi\rangle
=
-\frac{1}{2}\int
\tilde\psi^* \tilde{\nabla}^2\tilde\psi\,
d^3\tilde{r}.
\end{equation}

Using

\begin{equation}
\int
\tilde{\nabla}\cdot
(\tilde\psi^*\tilde{\nabla}\tilde\psi)\,
d^3\tilde{r}=0,
\end{equation}

we obtain

\begin{equation}
E_{\rm kin}
=
\frac{1}{2}\int
|\tilde{\nabla}\tilde\psi|^2\,
d^3\tilde{r}.
\end{equation}

Using

\begin{eqnarray}
|\tilde{\nabla}\tilde\psi|^2
&=&
\left(\frac{\tilde{r}}{\sigma^2}\right)^2
|\tilde\psi|^2,
\qquad
\langle \tilde{r}^2\rangle
=
\frac{3}{2}\sigma^2\equiv \tilde{R}_g^\mathcal{A},
\end{eqnarray}

we get

\begin{equation}
E_{\rm kin}
=
\frac{1}{2\sigma^4}
\int \tilde{r}^2|\tilde\psi|^2\,
d^3\tilde{r}
=
\frac{1}{2\sigma^4}
\langle \tilde{r}^2\rangle
\int |\tilde\psi|^2\,d^3\tilde{r}.
\end{equation}

\begin{equation}
\begin{aligned}
E_{\rm kin}
&=
\frac{1}{2\sigma^4}
\left[
\frac{3}{2}\sigma^2
\right]N
=
\boxed{\frac{3N}{4\sigma^2}}.
\end{aligned}
\end{equation}

Next, we calculate self-interaction energies:

\begin{equation}
E_g
=
\frac{\tilde g}{2}
\int |\tilde\psi|^4\,d^3\tilde{r},
\qquad
|\tilde\psi|^4
=
\frac{N^2}{\pi^3\sigma^6}
e^{-2\tilde{r}^2/\sigma^2}.
\end{equation}

Therefore,

\begin{eqnarray}
\int |\tilde\psi|^4\,d^3\tilde{r}
&=&
\frac{N^2}{\pi^3\sigma^6}
\int e^{-2r^2/\sigma^2}\,d^3\tilde{r}
\nonumber\\
&=&
\frac{N^2}{\pi^3\sigma^6}
\left(
\frac{\sqrt{\pi}\,\sigma}{\sqrt{2}}
\right)^3
\nonumber\\
&=&
\frac{N^2}
{2\sqrt{2}\,\pi^{3/2}\sigma^3}.
\end{eqnarray}

Hence,

\begin{equation}
E_g
=
\boxed{
\frac{\tilde g N^2}
{4\sqrt{2}\,\pi^{3/2}\sigma^3}
}.
\end{equation}

Similarly,
\begin{eqnarray}
\int |\tilde\psi|^6\,d^3\tilde{r}
&=&
\frac{N^3}{\pi^{9/2}\sigma^9}
\left(
\frac{\sqrt{\pi}\,\sigma}{\sqrt{3}}
\right)^3
\nonumber\\
&=&
\frac{N^3}
{3\sqrt{3}\,\pi^3\sigma^6}.
\end{eqnarray}
So
\begin{equation}
E_{g_2}
=
\frac{\tilde g_2}{3}
\int |\tilde\psi|^6\,d^3\tilde{r}
=
\boxed{
\frac{\tilde g_2N^3}
{9\sqrt{3}\,\pi^3\sigma^6}
}.
\end{equation}

Now for the gravitational energy, we use

\begin{equation}
\tilde\Phi
=
-\frac{\tilde G}{4\pi}
\int
\frac{\rho(\mathbf r')}
{|\mathbf r-\mathbf r'|}
d^3r'.
\end{equation}

\begin{eqnarray}
E_G&=&\frac{1}{2}\int \rho(\mathbf r)\,
\tilde\Phi(\mathbf r)\,d^3r\\
E_G&=&-\frac{\tilde G}{8\pi}
\int d^3\tilde{r}\,d^3\tilde{r}'
\frac{
|\tilde\psi(\tilde{\mathbf{r}})|^2
|\tilde\psi(\tilde{\mathbf{r}}')|^2
}{
|\tilde{\mathbf{r}}-\tilde{\mathbf{r}}'|
}.
\end{eqnarray}

Using the Fourier representation

\begin{equation}
\frac{1}{|\tilde{\mathbf{r}}-\tilde{\mathbf{r}}'|}
=
\int
\frac{d^3k}{(2\pi)^3}
\frac{4\pi}{k^2}
e^{i\mathbf k\cdot
(\tilde{\mathbf{r}}-\tilde{\mathbf{r}}')},
\end{equation}

we obtain

\begin{eqnarray}
E_G
&=&
-\frac{\tilde G}{8\pi}
\int d^3\tilde{r}
\int d^3\tilde{r}'
|\tilde\psi(\tilde{\mathbf{r}})|^2
|\tilde\psi(\tilde{\mathbf{r}}')|^2
\nonumber\\
&&\times
\int
\frac{d^3k}{(2\pi)^3}
\frac{4\pi}{k^2}
e^{i\mathbf k\cdot
(\tilde{\mathbf{r}}-\tilde{\mathbf{r}}')}.
\end{eqnarray}

Define

\begin{equation}
\tilde\psi(\mathbf k)
\equiv
\int d^3\tilde{r}\,
|\tilde\psi(\tilde{\mathbf{r}})|^2
e^{-i\mathbf k\cdot\tilde{\mathbf{r}}}.
\end{equation}

Since $|\tilde\psi(\tilde{\mathbf{r}})|^2$ is real,

\begin{equation}
\tilde\psi(-\mathbf k)
=
\tilde\psi^*(\mathbf k).
\end{equation}

Therefore,

\begin{equation}
E_G
=
-\frac{\tilde G}{8\pi}
\int
\frac{d^3k}{(2\pi)^3}
\frac{4\pi}{k^2}
|\tilde\psi(\mathbf k)|^2.
\label{eq:grav_E}
\end{equation}

Now,

\begin{equation}
|\tilde\psi(\tilde{\mathbf{r}})|^2
=
\frac{N}{\pi^{3/2}\sigma^3}
e^{-\tilde{r}^2/\sigma^2}.
\end{equation}

Thus,

\begin{equation}
\begin{aligned}
\tilde\psi(\mathbf k)
&=
\frac{N}{\pi^{3/2}\sigma^3}
\left[
\int dx\,
e^{-\frac{x^2}{\sigma^2}-ik_xx}
\right]
\\
&\quad\times
\left[
\int dy\,
e^{-\frac{y^2}{\sigma^2}-ik_yy}
\right]
\\
&\quad\times
\left[
\int dz\,
e^{-\frac{z^2}{\sigma^2}-ik_zz}
\right].
\end{aligned}
\end{equation}

Completing the square,

\begin{equation}
-\frac{x^2}{\sigma^2}-ik_xx
=
-\frac{1}{\sigma^2}
\left(
x+\frac{ik_x\sigma^2}{2}
\right)^2
-\frac{k_x^2\sigma^2}{4},
\end{equation}

we have

\begin{eqnarray}
\int dx\,
e^{-x^2/\sigma^2-ik_xx}
&=&
e^{-k_x^2\sigma^2/4}
\int du\,e^{-u^2/\sigma^2}
\nonumber\\
&=&
\sqrt{\pi}\,\sigma\,
e^{-k_x^2\sigma^2/4}.
\end{eqnarray}

Consequently,

\begin{eqnarray}
\tilde\psi(\mathbf k)
&=&
\frac{N}{\pi^{3/2}\sigma^3}
\left(
\pi^{3/2}\sigma^3
e^{-\frac{\sigma^2}{4}
(k_x^2+k_y^2+k_z^2)}
\right)
\nonumber\\
&=&
N e^{-k^2\sigma^2/4}.
\end{eqnarray}

Hence,

\begin{equation}
|\tilde\psi(\mathbf k)|^2
=
\boxed{
N^2e^{-k^2\sigma^2/2}
}.
\end{equation}

Substituting this into Eq.~\eqref{eq:grav_E}, we get

\begin{eqnarray}
E_G
&=&
-\frac{\tilde G}{8\pi}
\int_0^\infty
\frac{4\pi k^2\,dk}{(2\pi)^3}
\frac{4\pi}{k^2}
N^2e^{-k^2\sigma^2/2}
\nonumber\\
&=&
-\frac{\tilde G N^2}{4\pi^2}
\int_0^\infty
e^{-k^2\sigma^2/2}\,dk\\
E_G&=&
\boxed{
-\frac{\tilde G N^2}
{4\sqrt{2}\pi^{3/2}\,\sigma}
}.
\end{eqnarray}

Thus, adding all those energies gives us the total energy as:

\begin{equation}
E(\sigma)
=
\boxed{
\frac{3N}{4\sigma^2}
-\frac{\tilde G N^2}
{4\sqrt{2}\pi^{3/2}\,\sigma}
+\frac{\tilde gN^2}
{4\sqrt{2}\,\pi^{3/2}\sigma^3}
+\frac{\tilde g_2N^3}
{9\sqrt{3}\,\pi^3\sigma^6}
}.
\label{eq:energy_ansatz}
\end{equation}

A similar relation to Eq.~\eqref{eq:energy_ansatz} for cubic-GPPE has beein derived in Ref.~\cite{chavanis2011mass}. Substituting $\tilde g=\lambda\sqrt{\tilde g_2}$ and imposing the equilibrium condition $\frac{\partial E}{\partial\sigma}=0$ gives,

\begin{equation}
-\frac{3N}{2\sigma^3}
+
\frac{\tilde GN^2}
{4\sqrt{2}\pi^{3/2}\,\sigma^2}
-
\frac{3\lambda\sqrt{\tilde g_2}\,N^2}
{4\sqrt{2}\,\pi^{3/2}\sigma^4}
-
\frac{2\tilde g_2N^3}
{3\sqrt{3}\,\pi^3\sigma^7}
=0.
\label{eq:sigma_N_again}
\end{equation}

\end{document}